\documentclass[prl,twocolumn,amsmath,amssymb,superscriptaddress]{revtex4-1}

\usepackage{graphicx}
\usepackage{dcolumn}
\usepackage{hhline}
\usepackage{bm,color}
\usepackage{float}
\usepackage{amsmath}
\usepackage{amsfonts}
\usepackage{amssymb}
\usepackage{epstopdf}

\begin{document}

\title{Electronic structure and magnetism of transition metal dihalides: bulk to monolayer}

\author{A. S. Botana}
\email{antia.botana@asu.edu}
\affiliation{Materials Science Division, Argonne National Laboratory, Argonne, IL 60439}
\author{M. R. Norman}
\affiliation{Materials Science Division, Argonne National Laboratory, Argonne, IL 60439}

\begin{abstract}
Based on first principles calculations, the evolution of the electronic and magnetic properties of transition metal dihalides MX$_2$ (M= V, Mn, Fe, Co, Ni; X = Cl, Br, I) is analyzed from the bulk to the monolayer limit. A variety of magnetic ground states is obtained as a result of the competition between  direct exchange and superexchange. The results predict that FeX$_2$, NiX$_2$, CoCl$_2$ and
CoBr$_2$ monolayers are ferromagnetic insulators with sizable magnetocrystalline anisotropies. This makes them ideal candidates for robust ferromagnetism at the single layer level.
Our results highlight the importance of spin-orbit coupling to obtain the correct ground state.

\end{abstract}

\maketitle

\section{Introduction}

Long range magnetic order is a common phenomena in three-dimensional materials but not in lower dimensions. According to the
Mermin-Wagner theorem, long-range magnetic order is not possible in 2D for spin-rotational-invariant systems.\cite{mermin_wagner}  However, magnetic anisotropies (i.e., those that break spin-rotational symmetry) remove this restriction.  Magnetic van der Waals
(vdW) materials are good candidates given their flexibility that can allow for a tuning of their magnetic anisotropy, as well as for their ease of exfoliation. They offer the possibility of obtaining a magnetic ground state even in the single-layer limit.

An example is CrI$_3$, a layered Ising ferromagnet (FM) in which the Cr ions lie on a honeycomb lattice.
In 2017, it was shown that ferromagnetism does survive
at the single-layer level with a transition temperature near that of the bulk material (61 vs 45 K).\cite{cri3_nature}  
Magnetism in 2D has the potential to open up a number of technological opportunities such as sensing, information, and data storage. 2D ferromagnets are particularly interesting for spintronic applications.
Novel functionalities
based on van der Waals heterostructures, in which magnetism adds a new ingredient, can also be anticipated. The potential use of CrI$_3$ and other materials for
building devices and tuning their properties through gating has already started to be
explored, marking the birth of a new era of magnetism.\cite{cr2ge2te6_1, cri3_electric_field, cr2ge2te6, 2d_review}
However, pushing T$_c$ to higher values will be necessary for real applications.

\begin{figure}
\includegraphics[width=0.8\columnwidth]{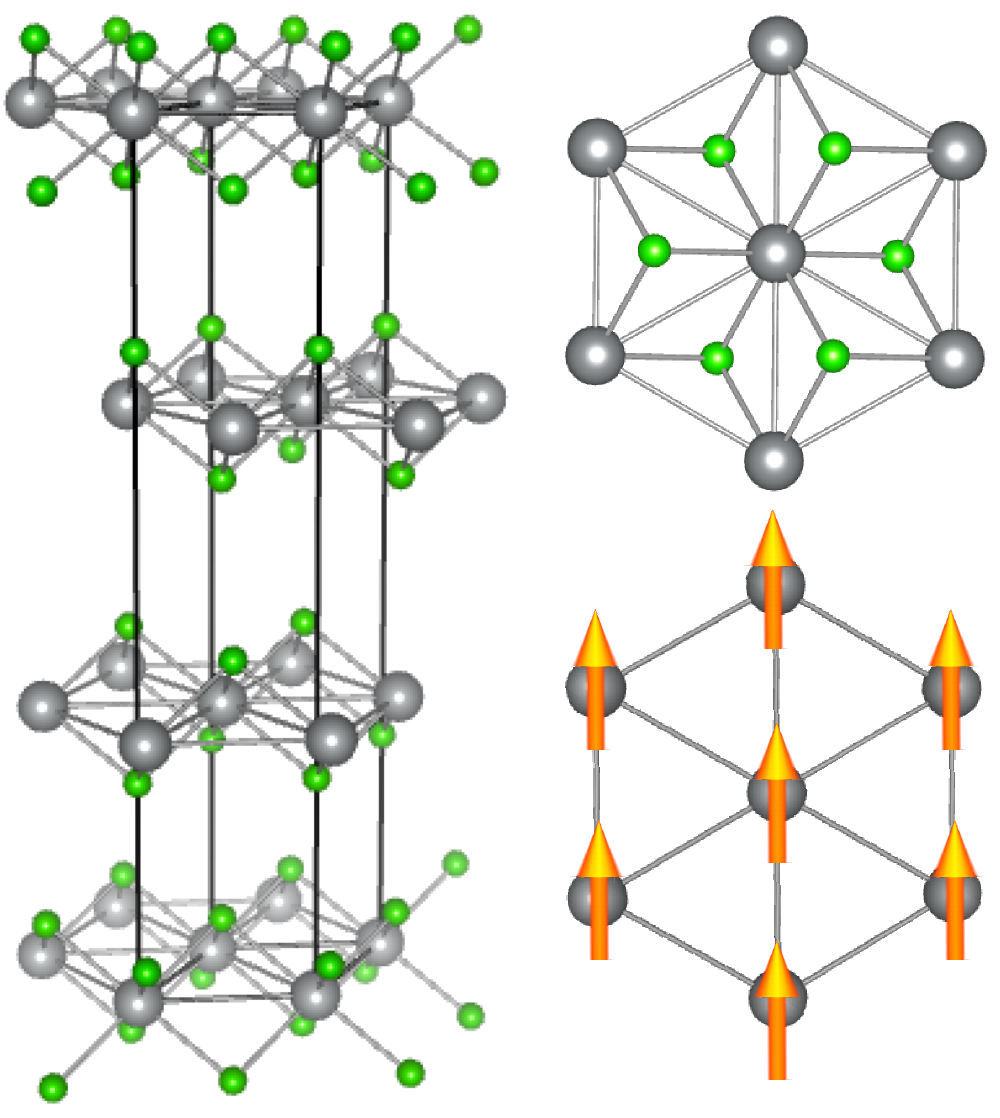}
\caption {Left panel: Crystal structure of transition metal dihalides (MX$_2$) in the 1-T phase showing the triangular lattice formed by the magnetic (M) atoms (in gray). Halide (X) atoms are shown in green. Right panel: Top view of the effective triangular ferromagnetic lattice formed by the metal atoms for Fe, Co, and Ni dihalide monolayers.}
\label{fig1}
\end{figure}

In this regard, there is a materials family related to CrI$_3$ that holds equal promise: binary transition metal dihalides MX$_2$ (M = transition metal, X = halogen: Cl, Br, I). They form low dimensional crystal structures composed of either one dimensional chains or two dimensional layers.\cite{exp_dihalides}  The layered structure is shown in Fig.~\ref{fig1} where the vdW gap between layers is apparent.  In contrast to trihalides, layered dihalides contain a triangular lattice of transition metal cations, and geometrical frustration in such a lattice is expected when the magnetic interactions are antiferromagnetic (AFM).
This set of materials hence provides a rich playground for examining low dimensional magnetism. In the bulk, the magnetism of most of these materials was analyzed decades ago.\cite{exp_dihalides}
At the monolayer level, some theoretical effort has been devoted to them,\cite{hennig, chlorides, dihalides_mono, v_dih_3, fecl2_hm}  but the full trends in magnetism and electronic structure and in particular the magnetocrystalline anisotropy energy (crucial to establish 2D long range magnetic order) have not been completely studied for all possible 3$d$ transition metal and halide ions.

Here, we present a systematic study of the electronic structure and magnetism of MX$_2$  compounds from the bulk to the monolayer limit by means of first principles calculations. After a general overview of the magnetic trends in bulk dihalides, we turn to results at the monolayer level, where a variety of magnetic states is found as a result of the competition between direct
exchange and superexchange via the halogen $p$ states.
We predict that FeX$_2$, NiX$_2$, CoCl$_2$ and
CoBr$_2$ monolayers are ferromagnetic insulators with an easy axis normal to the planes and a sizable magnetocrystalline anisotropy making them ideal candidates for robust 2D ferromagnetism. Our results also highlight the importance of considering both the trigonal distortion of the MX$_6$ octahedra as well as the spin-orbit coupling to obtain the correct ground state.

\section{Computational Methods}

Our electronic structure calculations were performed using the all-electron, full potential
code WIEN2k \cite{wien2k} based on the augmented plane wave
plus local orbitals (APW + lo) basis set.
The Perdew-Burke-Ernzerhof version of the generalized
gradient approximation (GGA) \cite{pbe} was used for structural relaxation and optimization. Then, the LDA+$U$ scheme within the fully localized limit was applied to treat the strong correlations.\cite{sic} We have studied the evolution of the electronic structure with increasing $U$  ($U$= 2-6 eV, $J$= 0.8 eV). For all materials, we performed calculations in supercells of size 1$\times$$\sqrt{3}$ to allow for the possibility of an in-plane antiferromagnetic `striped' configuration.

For the calculations, we converged using R$_{mt}$K$_{max}$ = 7.0 with
a fine $k$ mesh of 17$\times$17$\times$3 for the bulk materials and 19$\times$19$\times$1 for the monolayers. Muffin-tin radii of 2.48 a.u.~for Fe, 2.45 a.u.~for V, 2.45 a.u.~for Co, 2.50 a.u.~for Mn, 2.50 a.u.~for Ni, 2.11 a.u.~for Cl, 2.48 a.u.~for Br, and 2.50 a.u.~for I were employed.

To determine the magnetocrystalline anisotropy energy (MAE), we calculate
how the direction of the spin of the metal atom affects the energy
when spin-orbit coupling  (SOC) is included with the moment orientation either in-plane or out-of-plane.  SOC was introduced in a second variational procedure.\cite{soc} The calculations of magnetic anisotropy require careful
convergence of the total energy.
We found that a converging criterion for the total energy to within 10$^{-6}$ eV yields stable results.

\section{Structural properties}

As mentioned above, most of
the bulk transition metal dihalides have a natural
layered structure that contains triangular
nets of cations in edge-sharing octahedral coordination forming MX$_2$ layers separated
by van der Waals gaps (Fig.~\ref{fig1}).  The octahedral crystal field will split the
$3d$ orbitals of the metal atoms into higher lying e$_g$ ($d_{x^2-y^2}$, $d_{z^2}$) and lower lying t$_{2g}$
($d_{xy}$, $d_{xz}$, $d_{yz}$) states. All of the nearest-neighbor distances within the MX$_6$ octahedra are the same, but there is a trigonal distortion which further splits the t$_{2g}$ manifold. The consequences this has for the electronic structure are explained below.

MX$_2$ compounds adopt either the trigonal CdI$_2$ structure (so called 1-T with $P\bar3m1$ space group) or the rhombohedral CdCl$_2$ ($R\bar3m$) one. These structures have different stackings along the $c$-axis.
The Cr and Cu materials are slightly different in that their structure in the bulk is 1D-like and monoclinic, respectively, hence we do not analyze them here.
The structure of monolayer dihalides is analogous to that of the intensively studied transition metal dichalcogenides in which ferromagnetism at the monolayer level has been anticipated by DFT calculations.\cite{dichal_FM}
Regardless, based on the known bulk values, larger magnetic moments in 2D dihalides can be expected.

It was previously found that all dihalides prefer the 1-T crystal
structure at the monolayer level.\cite{dihalides_mono} The stability of
single-layer dihalides was also evaluated from their formation energy confirming that not only are metal dihalide monolayers stable, but also that they could potentially be exfoliated.\cite{marzari_exf, dihalides_mono}  Based on this, for the monolayer we will focus on 1-T structures only.
The calculated structural parameters of MX$_2$ monolayers are summarized
in Table \ref{table1} and agree with Ref.~\cite{dihalides_mono} and with the experimental bulk data.

\begin{table}
\caption{Calculated in-plane lattice parameters for monolayer MX$_2$ within GGA.}
\begin{ruledtabular}
\begin{tabular}{lccc}
\multicolumn{1}{l}{$a$ } &
\multicolumn{1}{l}{Cl} &
\multicolumn{1}{c}{Br}&
\multicolumn{1}{c}{I} \\

   \hline
   VX$_2$ & 3.62 & 3.81  & 4.08   \\ 
      MnX$_2$ & 3.64 & 3.84  & 4.12   \\ 
            FeX$_2$ & 3.49 & 3.69  & 3.98   \\ 
                  CoX$_2$ & 3.49 & 3.73  & 3.92   \\ 
                   NiX$_2$ & 3.45 & 3.64  & 3.92
\end{tabular}
\end{ruledtabular}
\label{table1}
\end{table}

\section{Dihalides - Bulk magnetism}

We analyze first the electronic structure and magnetism of the bulk materials, experimentally studied already, to test the validity of our predictions. 
All the materials are insulators as shown by experiment and confirmed by our calculations (to open up a gap, a $U$ is required in some cases). 
As mentioned above, there are two competing magnetic interactions in the planes: direct exchange between transition metal
cations, and superexchange through the halogen
anions. The magnetism of these materials can be understood from the
Goodenough-Kanamori-Anderson rules.\cite{goodenough} In the case of a 90$^\circ$ M-X-M bond angle, the e$_g$-e$_g$ exchange is always FM and weak, the
direct t$_{2g}$-t$_{2g}$ overlap can give rise to an AFM exchange, and depending on the particular
orbital occupation, the t$_{2g}$-t$_{2g}$ superexchange via halides can either be AFM or weakly FM.
For the case of edge-sharing octahedra, the t$_{2g}$ orbitals
on neighboring sites, pointing between the oxygens, are directed toward each
other. The resulting $d-d$ hopping turns out to
be very important for early $3d$ metals  (i.e., V) and it can give rise to AFM
exchange.
As a note, the magnetic order found in most of these compounds either consists on ferromagnetic planes, stripes,
or is helimagnetic.\cite{exp_dihalides} 
The results of the density functional theory (DFT) calculations are shown in Table \ref{table2}.  We discuss them below in order of increasing $d$ count.
Three magnetic structures were explored: AFS (stripe) order, which was taken to be alternating rows of ferromagnetic spins stacked AFM along $c$, AF order taken to be
ferromagnetic planes stacked AFM along $c$, and FM order. These allow us to capture the physics of most of the materials. However, the non-colinear nature of the order observed in some of the materials is beyond the scope of the present work.

\begin{table}
\caption{Magnetism of bulk MX$_2$ within GGA. The second column reflects the nature of the ground state (GS) where AFS stands for an antiferromagnetic striped
configuration (in-plane stripes coupled AFM along $c$), AF for FM planes coupled AFM along $c$, FM for ferromagnetic, I for insulator, and HM for half-metal. * indicates that FeX$_2$ become insulators after SOC and $U$ are included. The third column shows the energy difference (per unit cell) between ferromagnetic and corresponding antiferromagnetic ordering. The fourth column shows the magnetocrystalline anisotropy energy for the magnetic moments pointing in-plane versus out-of-plane (negative values reflect out-of-plane moments). The last column shows the calculated magnetic moment.}
\begin{ruledtabular}
\begin{tabular}{lcccc}
\multicolumn{1}{l}{} &
\multicolumn{1}{c}{GS} &
\multicolumn{1}{c}{$\Delta$E$_{FM-AFM}$ (meV)}&
\multicolumn{1}{c}{MAE (meV)} &
\multicolumn{1}{c}{MM}($\mu_B$)  \\

   Bulk &   &  & & \\ 
   \hline
   VCl$_2$ & AFS-I & 55.49 & 0.04 & 2.60  \\ 
           VBr$_2$ & AFS-I & 25.57  & 0.03 & 2.61  \\ 
                 VI$_2$ & AFS-I & 10.13 & 0.62 & 2.62  \\ 
   \hline     
      MnCl$_2$ & AFS-I & 18.45  & 0.09 & 4.50  \\ 
      MnBr$_2$ & AFS-I & 12.27 & 0.03 & 4.52  \\ 
           MnI$_2$ & AFS-I & 8.63 & 0.16 & 4.51  \\ 
   \hline     
      FeCl$_2$ & AF-HM*& 14.74  & -1.09 & 3.47 \\ 
      FeBr$_2$ & AF-HM* & 5.44  & -0.82 & 3.54  \\ 
           FeI$_2$ & AF-HM* & 2.72  & -1.05 & 3.40  \\      
    \hline     
      CoCl$_2$ & AF-I & 5.03  & 0.62 &  2.54\\ 
      CoBr$_2$ & AF-I & 1.49  & 0.76 &  2.49\\ 
           CoI$_2$ & AF-I & 13.60  & 0.17 & 2.24    \\
   \hline     
      NiCl$_2$ & AF-I & 4.62  & 0.50 & 1.46 \\ 
      NiBr$_2$ & AF-I & 2.99  & 0.26 & 1.39  \\ 
           NiI$_2$ & AF-I & 27.33  & 0.30 & 1.25

\end{tabular}
\end{ruledtabular}
\label{table2}
\end{table}

\textit{VX$_2$.} In these materials, V is in a high spin $d^3$ configuration (S = 3/2). 
Neutron powder diffraction studies found that all three vanadium dihalides order
antiferromagnetically with N\'{e}el temperatures of 36.0, 29.5, and 16.3 K for Cl, Br and I, respectively,\cite{v_dih_2} with the low values relative to the Weiss temperatures
(437 K, 335 K, 143 K) \cite{vx2_mag} being an indication of geometric frustration.
Our AFS (stripe) calculations indeed confirm that the magnetic order involves antiferromagnetic in-plane coupling, with the moments predicted to lie in-plane. The actual moment orientation for VCl$_2$ and VBr$_2$ is non-colinear (120 degree orientation of the moments) as typical for triangular antiferromagnets,
and moreover the moments appear to be tilted out of the plane.\cite{v_dih_2, v_dih_1, vcl2}.
VI$_2$ is more complicated in that the 120 degree order occurs first, and then slightly below this at 14.4 K a stripe phase (the same as simulated here) sets in,
though the moments are probably also tilted out of the plane.\cite{vi2,v_dih_2}
The closeness of the two phase transitions indicates that the free energy difference between the 120 degree order and the stripe phase is small.

\begin{figure*}
\includegraphics[width=1.98\columnwidth]{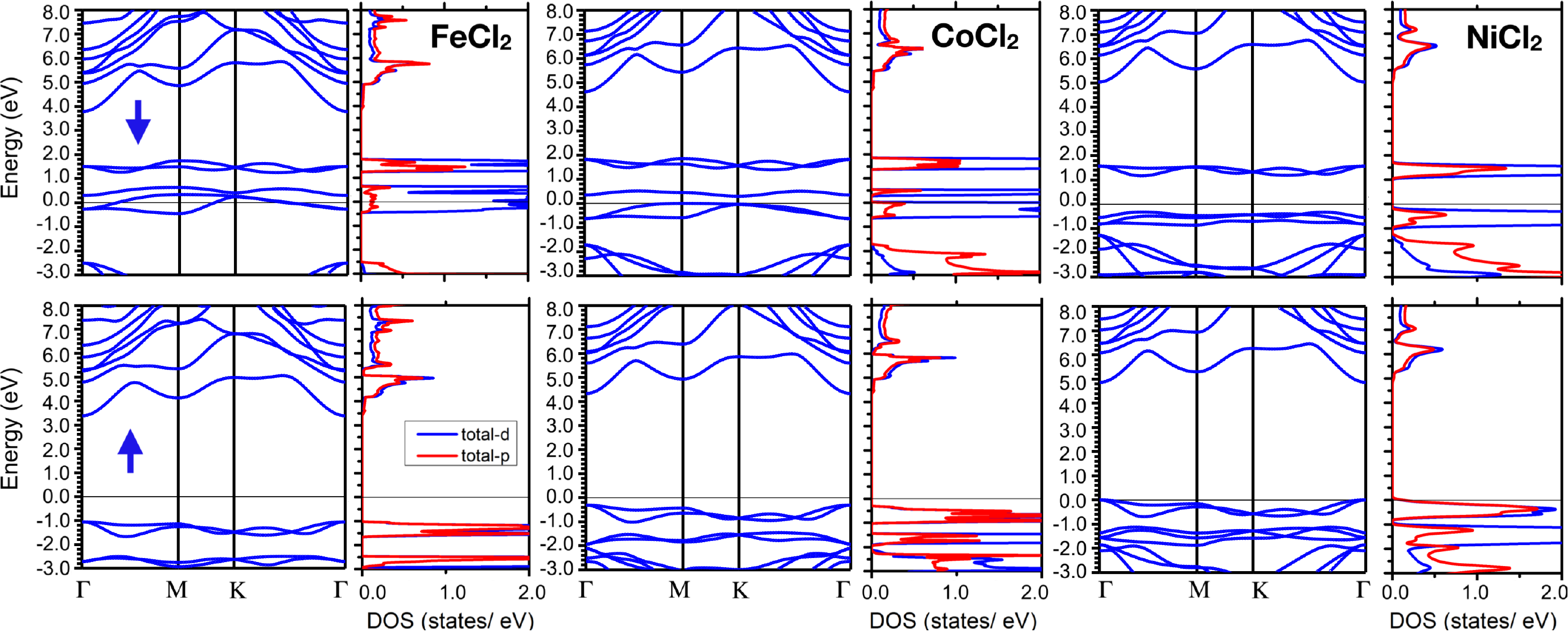}
\caption {Atom-resolved density of states (DOS) and band structures for different dihalide monolayers within GGA: FeCl$_2$, CoCl$_2$ and NiCl$_2$. Majority and minority spin channels are represented by up and down arrows.}
\label{fig2}
\end{figure*}

\textit{MnX$_2$.} In these materials, Mn is in a high spin $d^5$ configuration (S = 5/2). For this filling, superexchange is AFM. The magnetic structure for MnCl$_2$ has
ferromagnetic stripes of width two rows within the layers with antiferromagnetic coupling between neighboring stripes and between the layers.\cite{mn_dih_2} There is evidence that the moments lie in-plane for the Cl and Br materials,\cite{mnbr2_neutrons} that our DFT calculations confirm.
For simplicity, we approximate this state by the one width row AFS (stripe) order for the purposes of Table II. 
MnI$_2$ adopts a complicated helical magnetic structure at low T stemming from competition with longer range exchange.\cite{vi2_mag}
The development of a ferroelectric polarization was recently reported, spurring interest in this compound as a multiferroic material.\cite{multif}

\textit{FeX$_2$.} Divalent iron, $d^6$, is expected to be in a high spin state, S = 2. Given the partially filled $t_{2g}$ levels, an orbital moment may be expected as well. 
 The magnetic structure of FeCl$_2$ and FeBr$_2$ shows ferromagnetic order within the layers and antiferromagnetic stacking along $c$.\cite{fe_cl_br}
 Our calculations do reproduce the observed magnetic order. 
 The iodide counterpart, though, adopts the two-row width stripes as found in MnBr$_2$.\cite{fei2}
 The moments are along the $c$ axis in all cases. Our obtained magnetocrystalline anisotropy energies confirm that the moments lie out-of-plane.

\textit{CoX$_2$.} Co$^{2+}$, $d^7$, can be in a high spin (S = 3/2) or
low spin (S = 1/2) state. Neutron diffraction results show that the high spin state is preferred, at least for CoCl$_2$ and CoBr$_2$, and our DFT calculations confirm this picture.\cite{fe_cl_br}
Below their ordering temperatures, both of these compounds adopt a magnetic structure
 with ferromagnetic alignment within each layer and antiferromagnetic stacking. This is correctly reproduced by our calculations, along with the fact that the magnetic moments lie in-plane.
The magnetic behavior in CoI$_2$ is more complex - it is a helimagnet with a spiral spin structure, indicating the presence of longer range exchange.\cite{coi2}

\textit{NiX$_2$.} Divalent nickel in these compounds has a $d^8$ configuration, with low spin being S=0 and high spin S=1, the later being favored.  Both NiCl$_2$ and NiBr$_2$
show a magnetic configuration with moments lying within the plane that are ferromagnetically aligned within each layer, with
antiferromagnetic stacking.\cite{nibr2} Our DFT calculations are able to reproduce this. NiBr$_2$ develops helimagnetic order below a lower second transition, that is seen as well below T$_N$ in the iodide material. Like MnI$_2$ and CoI$_2$ described above, NiBr$_2$ and NiI$_2$ also develop a ferroelectric
polarization in their helimagnetic states. \cite{multif_2, multif_3}

The trend in the magnetic moments across the MX$_2$ series is the expected one according to the spin states described above: starting with
$\sim$3 $\mu_B$ per metal atom for the V-halides, increasing to $\sim$5 $\mu_B$ for the Mn-halides, and
then gradually decreasing to $\sim$2 $\mu_B$ for the Ni-halides  (Table \ref{table2}). The halogens develop a
magnetic moment of 0.16-0.20 $\mu_B$ and hence show a significant spin-polarization.The magnetic moment on the metal atoms decreases with the Z of the halide atom (Cl - Br - I). In the Fe and Co compounds, an orbital moment $\sim$ 0.1 and  0.2 $\mu_B$, respectively, is found when SOC is included. 

\section{dihalides-monolayer magnetism}

After confirming the correct magnetic ordering trends can be reproduced in the bulk dihalides, we turn our attention to the monolayers. The obtained ground states are very similar to those obtained in the bulk, a reasonable outcome given the weak interlayer coupling in vdW materials.
In a similar fashion to the above bulk description, comparison has been established between  ferromagnetic,
and antiferromagnetic (AFS stripe) ordering within the layers.
FeX$_2$, CoX$_2$ (X= Cl, Br) and NiX$_2$ prefer a FM ground state, while VX$_2$, 
MnX$_2$ and CoI$_2$ are (striped) AFS (Table \ref{table3}). These preferences
can be justified based on the above described competition between
 direct exchange and
superexchange for different fillings. CoI$_2$ breaks the trend among the Co compounds due to the sensitivity of superexchange for M-X-M bond angles near 90$^\circ$ to the relative position of the anion $p$ states (Ref.~\cite{goodenough}).

\begin{table}
\caption{Magnetism for monolayer MX$_2$ within GGA. The second column reflects the nature of the ground state where AFS stands for antiferromagnetic (stripe) in-plane order, FM for ferromagnetic in-plane order, I for insulator, and HM for half-metal. * indicates that FeX$_2$ become insulators after SOC and $U$ are included. The third column shows the energy difference between ferromagnetic and antiferromagnetic in-plane order. The fourth column shows the magnetocrystalline anisotropy energy for the magnetic moments pointing in-plane versus out-of-plane
(negative values reflect out-of-plane moments). The last column shows the calculated magnetic moment.}
\begin{ruledtabular}
\begin{tabular}{lcccc}
\multicolumn{1}{l}{} &
\multicolumn{1}{c}{GS} &
\multicolumn{1}{c}{$\Delta$E$_{FM-AFS}$ (meV)}&
\multicolumn{1}{c}{MAE (meV)} &
\multicolumn{1}{c}{MM}($\mu_B$)  \\

   Mono &   &  & & \\ 
   \hline
   VCl$_2$ & AFS-I & 61.47  & -0.53 & 2.67  \\ 
           VBr$_2$ & AFS-I & 31.90  & -0.30 & 2.67  \\ 
                 VI$_2$ & AFS-I & 10.70 & -0.25 & 2.67  \\ 
   \hline     
      MnCl$_2$ & AFS-I & 30.00  & -0.20 & 4.54  \\ 
      MnBr$_2$ & AFS-I & 15.36 & -0.25 & 4.52  \\ 
           MnI$_2$ & AFS-I & 12.92  & -0.18 & 4.46  \\ 
   \hline     
      FeCl$_2$ & FM-HM*& -121.58 & -0.89 & 3.57 \\ 
      FeBr$_2$ & FM-HM*& -67.66 & -0.33 & 3.53  \\ 
           FeI$_2$ & FM-HM*& -35.90  & -0.59 & 3.45  \\      
   \hline     
      CoCl$_2$ & FM-I & -57.52  & -0.69 & 2.54 \\ 
      CoBr$_2$ & FM-I & -10.33  & -0.68 & 2.49  \\ 
           CoI$_2$ & AFS-I & 14.15 & -0.50 & 2.23  \\       
              \hline     
      NiCl$_2$ &  FM-I & -38.76 & -0.12  & 1.68 \\ 
      NiBr$_2$ &  FM-I & -32.70  & -0.02 & 1.63  \\ 
           NiI$_2$ & FM-I & -33.73  & -0.18 & 1.53 

\end{tabular}
\end{ruledtabular}
\label{table3}
\end{table}

We will focus on the description of materials with a FM ground state at the monolayer level starting with the Fe compounds.
FeX$_2$ monolayers were the focus of previous studies in which they were predicted to be FM-half metals.\cite{hennig}
We have analyzed the electronic structure of these materials in further detail, in particular, the evolution of the electronic structure with $U$ within the LDA+$U$ method and upon inclusion of SOC. Fig.~\ref{fig2} shows the band structure and density of states of the half-metallic FeCl$_2$ monolayer obtained within GGA. 
The electronic structure is consistent with the description in Ref.~\onlinecite{hennig}: Fe$^{2+}$ being HS, the majority spin channel $d$ states are completely occupied; in the minority spin channel, there are two partially occupied t$_{2g}$ bands and three unoccupied $d$ bands (a flat t$_{2g}$ one, and above this, two $e_g$ ones). The electronic structure for the minority spin channel is identical to the bulk electronic structure in which FM layers are stacked in an AFM fashion (not shown). FeBr$_2$ and FeI$_2$ display the same basic electronic structure around E$_F$.  The spin moment per unit cell is 4$\mu_B$ with most of it residing on the Fe
site ($\sim$3.5$\mu_B$). 

\begin{figure}
\includegraphics[width=\columnwidth]{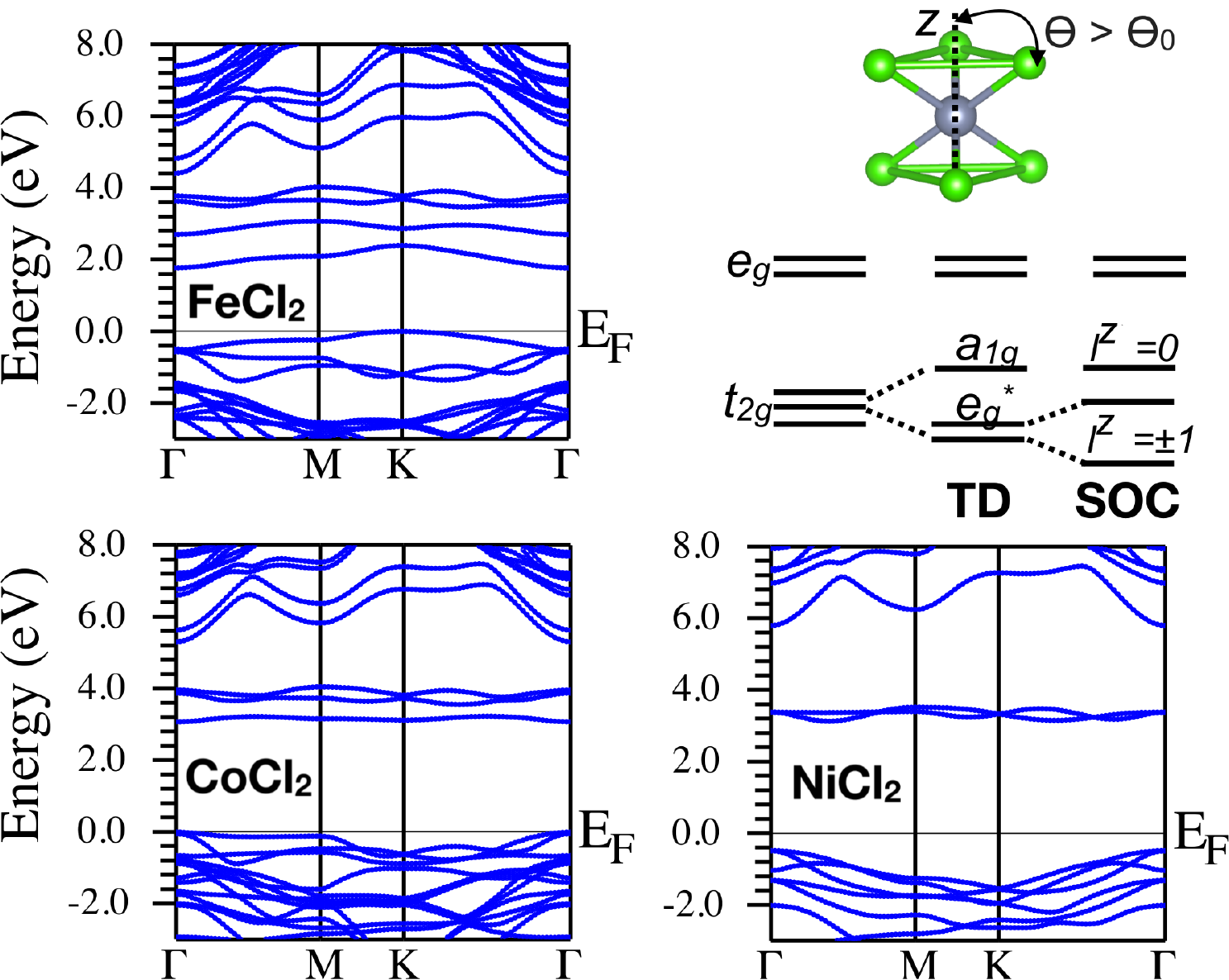}
\caption {Band structures for different dihalide monolayers within LDA+SOC+U (U= 4 eV): FeCl$_2$, CoCl$_2$ and NiCl$_2$. All of them are FM insulators. The top right panel shows the magnetic (M) atom surrounded by a trigonally distorted octahedron of halides. Distortions are determined by the angle $\theta$; the value cos$\theta_0$ = 1/$\sqrt3$ corresponds to an undistorted octahedron. The corresponding crystal field splitting of the $d$-orbitals of the magnetic atom is shown. The splitting of the t$_{2g}$ levels into $a_{1g}$ and $e_g$$^{*}$ is due to both the trigonal distortion (TD) of the octahedron and the contribution from neighboring M atoms to the crystal field. SOC additionally splits  the $e_g$$^{*}$ doublet.}
\label{fig3}
\end{figure}

One interesting question is the origin of the t$_{2g}$ splitting into an e$_g^{*}$ doublet and a higher-lying a$_{1g}$ singlet (Fig.~\ref {fig3}).\cite{kunes} This is due to the above mentioned trigonal distortion (compression along the [111] axis of the octahedra, i.e., the c-axis of the crystal). This distortion is quantified by the angle $\theta$ (Fig.~\ref{fig3}) that is larger than the value for an undistorted octahedron $\theta_0$= 54.74$^\circ$ = arccos(1/$\sqrt3$). Additionally, one has to take into account the contribution to the crystal field of neighboring transition metal atoms to the a$_{1g}$-e$_g$ splitting. The corresponding wavefunctions for these states can be written in the form:
\begin{equation}
\begin{aligned}
\begin{split}
\mid{a_{1g}}\rangle = \frac{1}{\sqrt{3}}\left(\mid{xy}\rangle+\mid{xz}\rangle+ \mid{yz}\rangle \right), \\[1pt]
\mid{e_{g \pm}^{*}}\rangle =\pm \frac{1}{\sqrt{3}}\left(\mid{xy}\rangle+e^{\pm2\pi i/3}\mid{xz}\rangle+e^{\mp2\pi i/3} \mid{yz}\rangle \right) 
\end{split}
\end{aligned}
\end{equation}

The a$_{1g}$ orbital has a very simple shape in local coordinates. It is analogous to a $z^2$- e$_g$ orbital with  its $z$ axis directed along the [111] axis (in this case, the c-axis of the crystal). The other two $t_{2g}$ orbitals, denoted as e$_g^{*}$, have a more complicated shape. These are states with $\mid{l^{z}}= \pm1 \rangle$ (with the quantization axis along c), the a$_{1g}$ state being the $\mid{l^{z}}= 0\rangle$ one. 

This clearly explains why, once SOC is included, the e$_g^{*}$ orbitals are split. The overall band structure is similar to that of Fig.~\ref{fig2} with an orbital  moment of 0.10 $\mu_B$ being induced on Fe. This splitting increases once an on-site $U$ is included, with an insulating gap opening up once $U$ exceeds 3 eV (Fig.~\ref{fig3}).  As a consequence, a larger orbital moment ($\sim$0.6 $\mu_B$) develops along the the trigonal axis, parallel to the spin moment. It should be noted that regardless of the $U$ value, a gap is not opened up unless SOC is included and the e$_g^{*}$ states are split.  The degeneracy lifting due to spin orbit coupling is similarly required to open a gap in FeX$_2$ bulk materials, which are known to be Mott insulators.

The electronic structure of monolayer CoCl$_2$ within GGA (d$^7$ filling, HS) is shown in Fig.~\ref{fig2}. The electronic structure for the bulk (with a magnetic order consisting of FM layers stacked AFM) is equivalent to that
of the monolayer. At the GGA level, a gap already opens up in spite of the partial ($d^7$) filling. The reason is once again the trigonal distortion of the octahedra that gives rise to the t$_{2g}$ splitting into an a$_{1g}$ singlet and e$_g^{*}$ doublet (the flat band right above the Fermi level in the minority spin channel has a$_{1g}$ character, the two e$_g^{*}$ are occupied). 
In a similar fashion to the Fe case, the total moment per unit cell is 3$\mu_B$ with most of it residing on the Co site
site ($\sim$2.5$\mu_B$).  Once a $U$ is included, the gap between the occupied e$_g^{*}$  and unoccupied  a$_{1g}$ states increases (Fig.~\ref{fig3}). The orbital moment derived for LDA+SOC+U (U= 4 eV) is 0.2 $\mu_B$.  This, unlike for the Fe case, is the same as for $U$=0, an expected result given the complete filling of the e$_g^{*}$ doublet.

NiCl$_2$ also is a FM insulator at the monolayer level. The insulating character with Ni being d$^8$ (HS) is simpler to understand. The band gap opens up between t$_{2g}$ occupied and e$_g$ unoccupied minority spin bands. The derived magnetic moment agrees with this picture (Table \ref{table3}). In this case, a $U$ simply increases the gap as expected (Fig.~\ref{fig3}). 

As mentioned above, the character of the ferromagnetic ordering in these 2D materials is determined by the magnetocrystalline
anisotropy energy. It is magnetic 2D materials with an easy
axis that can have a ferromagnetically ordered
phase at finite temperature. All Fe and Ni dihalides, as well as CoCl$_2$ and CoBr$_2$, are predicted to have their easy
axis along $c$. This makes them a very promising platform for stable and robust ferromagnetism.

As in Ref.~\onlinecite{dihalides_mono}, we estimate the strength of the magnetic
interactions from the energy difference between a FM and an AFM configuration (denoted as $\Delta$E).  In our case, for the
AFS state, a given magnetic ion has four AFM and two FM near neighbors, and for the FM state, six FM near neighbors.  The same counting applies for next near neighbors.
This leads to an effective $J$ of:
 \begin{align}
J = \frac{\Delta E}{8S^2}
 \end{align}
where $S$ is the spin per metal atom (2, 3/2 and 1 for Fe, Co and Ni). The calculated values for materials with a FM ground state are shown in Table \ref{table4}.
The derived values are similar to those in previous works in which pseudopotentials were used,\cite{dihalides_mono} with predicted values
for FeCl$_2$ and NiX$_2$ being particularly high,
noting that our estimates are higher than theirs since they assumed a factor of 12 rather than 8 in the denominator of Eq.~(2).
This $J$, which can be considered as a sum of the near-neighbor $J_1$ and next near-neighbor $J_2$, is relevant if longer range $J$s are not important (the next next near-neighbor $J_3$ drops out of this energy difference).  As a cautionary note, the values we estimate for $J$ typically exceed those extracted from inelastic
neutron scattering.\cite{birgeneau}  But, they should give some idea of the trends in $J$ as M and X are varied.

\begin{table}
\caption{Magnetism for monolayer MX$_2$ (continued). The second column reflects the nature of the ground state where AFS stands for antiferromagnetic (stripe) order, FM for ferromagnetic order, I for insulator, and HM for half-metal. * indicates that FeX$_2$ become insulators after SOC and $U$ are included. The third column shows the value of the exchange interaction. The fourth column is an estimate of the Curie temperature.}
\begin{ruledtabular}
\begin{tabular}{lccc}
\multicolumn{1}{l}{} &
\multicolumn{1}{c}{GS} &
\multicolumn{1}{c}{$J$ (K)}  &
\multicolumn{1}{c}{T$_c$}(K)  \\

   Mono &   &  &  \\ 
   \hline     
      FeCl$_2$ & FM-HM* &  44.1 &  160\\ 
      FeBr$_2$ & FM-HM* & 24.5 & 89\\ 
           FeI$_2$ & FM-HM* & 13.0  &  47 \\      
   \hline     
      CoCl$_2$ & FM-I & 37.1  & 135  \\ 
      CoBr$_2$ & FM-I & 6.7   &  24\\ 
           CoI$_2$ & AFS-I & --   & -- \\       
              \hline     
      NiCl$_2$ &  FM-I & 56.2  & 205 \\ 
      NiBr$_2$ &  FM-I & 47.4   & 173\\ 
           NiI$_2$ & FM-I & 48.9 &    178       

\end{tabular}
\end{ruledtabular}
\label{table4}
\end{table}

For uniaxial anisotropy, the appropriate model is the Ising one. For a triangular 2D lattice, the critical temperature is 3.641$J$/$k_B$.\cite{ising_1, ising_2}
The T$_c$s based on this, from our estimate of $J$, are shown in Table \ref{table4} for each of the monolayers.
Given the above caveats, these should be considered as overestimates.
Still, we would like to point out that in most cases, MX$_2$ materials have N\'{e}el temperatures that exceed those of their MX$_3$ counterparts.\cite{exp_dihalides}
Moreover, it is entirely conceivable that the monolayer T$_c$ could exceed the bulk T$_N$.  This
is particularly relevant for NiX$_2$, which have the highest T$_N$ of the MX$_2$ materials besides TiCl$_2$.\cite{exp_dihalides}
If the spins do convert from xy-like (in-plane) in the bulk to Ising-like (along $c$) in the monolayer, as we predict, then the NiX$_2$ monolayer Curie temperatures
could indeed be high.  As suggested in Ref.~\onlinecite{dihalides_mono}, T$_c$ could also be increased by strain from a suitable substrate.

\section{summary}

In summary, motivated by recent experiments for CrI$_3$, we have explored the evolution of the magnetism in transition metal dihalides from the bulk to the monolayer limit. FeX$_2$, NiX$_2$, CoCl$_2$ and CoBr$_2$ monolayers are predicted to be ferromagnetic insulators with out-of-plane moments and sizable magnetocrystalline anisotropies. 
Our results highlight the importance of considering the symmetry lowering at the transition metal site due to the trigonal distortion of the MX$_6$ octahedron, as well as including the effect of spin-orbit coupling, to obtain the correct ground states. This  work confirms the potential for stable and robust 2D ferromagnetism in transition metal dihalide-based monolayers, and we hope this will stimulate experimental efforts to realize them in the laboratory. 

\section{acknowledgments}
This work was supported by the  U.S.~Dept.~of Energy, Office of Science, Basic Energy Sciences, Materials Sciences and Engineering
Division. We acknowledge the computing resources provided on Blues, a high-performance computing cluster operated by Argonne's Laboratory Computing Resource Center.

\end{document}